\newif\ifsingle

\newif\ifFullversion
\Fullversiontrue 

\ifsingle
\documentclass[11pt,draftclsnofoot, onecolumn]{IEEEtran}		
\else		
\documentclass{article}
\usepackage{spconf}
\fi

\usepackage{times}
\usepackage{amsmath,dsfont}
\usepackage{amssymb,amsthm}
\usepackage{epsfig,verbatim}
\usepackage{subcaption}
\usepackage{setspace}
\usepackage{color}
\usepackage{cite}
\usepackage{epstopdf}
\usepackage{graphics}
\usepackage{accents}
\usepackage{acronym}
\usepackage[bookmarks,colorlinks]{hyperref}
\usepackage{booktabs}
\usepackage{mathtools}
\usepackage{enumitem}
\usepackage{multirow}
\usepackage{booktabs}

\usepackage[ruled,linesnumbered]{algorithm2e}  
\SetKwInput{KwData}{\textbf{Init}} 
\usepackage{algpseudocode} 
\newcommand{\myVec}[1]{{\boldsymbol{#1}}}

\newcommand{\mySet}[1]{\mathcal{#1}}
	
\newcommand{\Cnst}{M}
\newcommand{\CnstSet}{\mySet{M}}
\newcommand{\Nusers}{U}
\newcommand{\NusersSet}{\mySet{U}}
\newcommand{\Ntraining}{N_{T}}
\newcommand{\Ndata}{T_{\rm s}}
\newcommand{\Tdata}{t_{\rm d}}
\newcommand{\SGDIter}{\tau}

\newcommand{\FadeNet}{FedRec}

\definecolor{NewColor}{rgb}{0,0,0} 

\ifsingle

\else

\fi 

\acrodef{adc}[ADC]{analog-to-digital convertor}
\acrodef{cs}[CS]{compressed sensing}
\acrodef{dtft}[DTFT]{discrete-time Fourier transform}
\acrodef{dnn}[NN]{neural network} 
\acrodef{csi}[CSI]{channel state information}
\acrodef{map}[MAP]{maximum a-posteriori probability}
\acrodef{snr}[SNR]{signal-to-noise ratio}
\acrodef{bs}[BS]{base station} 
\acrodef{iot}[IOT]{Interent of Things}
\acrodef{mimo}[MIMO]{multiple-input multiple-output}
\acrodef{mse}[MSE]{mean-squared error}
\acrodef{pdf}[PDF]{probability density function}
\acrodef{rv}[RV]{random variable}
\acrodef{ml}[ML]{machine learning}
\acrodef{mf}[MF]{matched filter}
\acrodef{fec}[FEC]{forward error correction}
\acrodef{rs}[RS]{Reed-Solomon}
\acrodef{lti}[LTI]{linear time-invariant}
\acrodef{wss}[WSS]{wide-sense stationary}
\acrodef{psd}[PSD]{power spectral density}
\acrodef{ser}[SER]{symbol error rate} 
\acrodef{ber}[BER]{bit error rate} 
\acrodef{sgd}[SGD]{stochastic gradient descent} 
\acrodef{isi}[ISI]{intersymbol interference}  
\acrodef{awgn}[AWGN]{additive white Gaussian noise} 
\acrodef{ut}[UT]{user terminal} 
\acrodef{mmw}[mmWave]{millimeter wave}
\acrodef{noma}[NOMA]{non-orthognal multiple access}
\acrodef{mac}[MAC]{mulitple access channel}
\acrodef{fl}[FL]{federated learning}
\acrodef{ct}[CT]{continuous-time}

\title{FedRec: Federated Learning of Universal Receivers over Fading Channels}

	\name{
		{Mahdi Boloursaz Mashhadi, Nir Shlezinger, Yonina C. Eldar and Deniz G{\"u}nd{\"u}z
		} 
			\thanks{
		 D. G{\"u}nd{\"u}z received funding from the European Research Council (ERC) through project BEACON under grant No. 677854.	  Y. C. Eldar received funding from the European Union’s Horizon 2020 research and innovation program under grant No. 646804-ERC-COG-BNYQ,  and from the Israel Science Foundation under grant No. 0100101.
		M. B. Mashhadi and D. G{\"u}nd{\"u}z are with the Dept. of EE, Imperial College, London, UK (email: \{m.boloursaz-mashhadi, d.gunduz\}@imperial.ac.uk). 
		N. Shlezinger  is with the School of ECE, Ben-Gurion University of the Negev, Be'er-Sheva, Israel  (e-mail: nirshl@bgu.ac.il).   
		Y. C. Eldar is with the Faculty of Math and CS, Weizmann Institute, Rehovot, Israel (e-mail: yonina@weizmann.ac.il). 
		}
	\vspace{-1.0cm}
	}

\vspace{-0.75cm}

\begin{document}
\ninept            
	
	\maketitle
	\pagestyle{empty}
	\thispagestyle{empty}
	
\begin{abstract}
	Wireless communications is often subject to channel fading. Various statistical models have been proposed to capture the inherent randomness in fading, and conventional model-based receiver designs rely on accurate knowledge of this underlying distribution, which, in practice, may be complex and intractable. In this work, we propose a neural network-based symbol detection technique for downlink fading channels, which is based on the maximum a-posteriori probability (MAP) detector. To enable training on a diverse ensemble of fading realizations, we propose a federated training scheme, in which multiple users collaborate to jointly learn a universal data-driven detector, hence the name FedRec. The performance of the resulting receiver is shown to approach the MAP performance in diverse channel conditions without requiring knowledge of the fading statistics, while inducing a substantially reduced communication overhead in its training procedure compared to centralized training. 
	
	{}
\end{abstract}

\vspace{-0.2cm}
\section{Introduction}\label{sec:intro}
\vspace{-0.1cm}
Fading in wireless communications encapsulates the fact that the relationship between the transmitted signal and the received one is determined by the propagation of electromagnetic waves, which is typically dynamic and subject to different forms of randomness induced by the environment. Various distributions have been proposed to represent the statistical behavior of fading channels, including the Rayleigh, Rice, and Nakagami-$m$ models \cite{biglieri1998fading}, where each approximates the  propagation profile in different settings \cite{patzold2001mobile}.

The inherent randomness of fading channels makes symbol detection a challenging task. The common strategy is to  periodically transmit a-priori known pilot signals for the receiver to estimate the channel, which in turn is utilized for detection \cite{hassibi2003much}. The main drawback of this approach is that pilots must be transmitted anew each time the channel  changes, i.e., on each coherence duration, inducing notable overhead in rapidly-changing channels. Alternatively in fast fading conditions, one can utilize a single detection rule for all channel conditions, which accounts for its statistical model \cite{simon1998unified}. However, such model-based detection relies on the knowledge of the fading distribution, and tends to be inaccurate when the assumed distribution does not faithfully capture the real statistical propagation profile, or in the presence of a model mismatch.

An alternative strategy, which does not require the knowledge of the underlying statistical model, is to learn the detection mapping from data. In particular, \acp{dnn} have demonstrated unprecedented success over recent years in learning complex mappings in a data-driven fashion \cite{bengio2009learning}. Consequently, a multitude of \ac{dnn}-based receiver architectures that can operate without the prior knowledge of the underlying statistical model have been proposed, see, e.g., \cite{oshea2017introduction, simeone2018very, mao2018deep, gunduz2019machine, Pruning, balatsoukas2019deep}. \ac{dnn}-based receivers require labeled data to learn their mapping. If one has prior knowledge of the  channel conditions and can generate such data artificially, the \ac{dnn} can be trained offline. However, this may not be the case in many practical scenarios, where labeled data must be obtained from pilot transmissions. 

When the channel conditions change rapidly, NN-based receivers must be frequently retrained with new pilot signals, inducing significant computation and communication overhead. \ac{dnn}-based receivers can track dynamic channel conditions by periodic online training combined with methods to reduce the training complexity as in \cite{shlezinger2019viterbinet,shlezinger2019deepSIC,shlezinger2020data,park2020meta,teng2020syndrome}. Alternatively, one can train a single \ac{dnn} that would work for a broad range of channel conditions, by learning a universal rule based on the fading distribution rather than its specific realization \cite{farsad2018neural,liao2019deep}. Nonetheless, for a \ac{dnn} to learn the subtleties of a fading distribution, which may be complex and  intractable, the training data must contain a sufficiently large number of channel realizations. This may be difficult to achieve even with long pilot sequences, motivating the design of universal \ac{dnn}-based detectors for fading channels with limited pilots.

In this work, we propose \FadeNet, which is a data-driven universal symbol detection scheme for multi-user downlink fading channels, designed to learn its mappings from a limited amount of pilots. \FadeNet~is comprised of two algorithmic components: The first is the  \ac{dnn}-based symbol detection architecture, which uses sufficient statistics from the \ac{map} symbol detection rule as input features. This allows \FadeNet~to utilize a relatively compact \ac{dnn}, which can be trained with a smaller number of samples. The second component is the training mechanism, which exploits the fact that, in a wireless network of many users, while each user observes only a limited number of channel realizations, the realizations observed by the overall network are expected to be sufficiently diverse. \FadeNet~builds upon this insight  to have the users collaborate for training via \ac{fl} \cite{mcmahan2016communication,li2020federated,gunduz2020communicate}, allowing to train a single \ac{dnn} over a diverse dataset without additional pilots, at the cost of several iterations of parameter exchanges with the \ac{bs}. Our numerical results show that  \FadeNet~yields an accurate symbol detector, with a performance approaching that of a \ac{map} detector, and outperforms the model-based approach in the presence of inaccurate knowledge of the fading distribution. Moreover, \FadeNet~induces substantially less communication overhead compared to learning a \ac{dnn}-based symbol detector in a centralized fashion.

The rest of this paper is organized as follows: 
Section~\ref{sec:Model}  presents the system model and the problem formulation.
Section~\ref{sec:UnFadeNet} details the proposed \FadeNet~receiver. 
Numerical examples are presented in Section~\ref{sec:sims}. Finally, Section~\ref{sec:Conclusions}  concludes the paper. FedRec codes are available at: \url{https://github.com/MahdiBoloursazMashhadi/FedRec}


\vspace{-0.2cm}
\section{System Model}\label{sec:Model}
\vspace{-0.1cm}

We consider a downlink communication scenario, where a BS serves $\Nusers$ users indexed by $u \in \NusersSet \triangleq \{1,\ldots,\Nusers\}$. Although we focus on  single-antenna terminals in this paper, our approach can be extended to multiple-antenna terminals. Letting $x(t)\in \mySet{C}$ be the baseband \ac{ct} channel input transmitted by the \ac{bs} at time instance $t$, the corresponding channel output at the $u$th user is 
\begin{equation}
\label{eqn:ChModel}
r_u(t) = h_u(t)x(t) + w_u(t),\quad  u \in \NusersSet, 
\end{equation}
where $\{w_u(t)\}$ are independent identically distributed (i.i.d.) Gaussian noise signals with unit \ac{psd}, while $\{h_u(t)\}$ are i.i.d. flat fading coefficients, following a common distribution $p_h(\cdot)$. Various different models exist for $p_h(\cdot)$, which approximate the statistical behavior under different  channel conditions \cite[Ch. 2]{simon2005digital}, but we do not assume prior knowledge of $p_h(\cdot)$. 

Downlink communication is typically comprised of pilot and data transmission. During downlink data transmission to user $u$ commencing at time instance $\Tdata$, the \ac{bs} encodes the message $s_u$, uniformly distributed over $\CnstSet \triangleq \{1,\ldots, M\}$, into a signal of temporal duration of $\Ndata$ seconds, denoted by $x_{s_u}(t)$, $t\in \Tdata+[0,\Ndata)$. Each user $u\in\NusersSet$ uses its  channel output $r_u(t)$, obtained via \eqref{eqn:ChModel}, to recover $s_u$. During pilot transmission commencing at $t=0$, the \ac{bs} broadcasts a sequence of $\Ntraining$ a-priori known pilot symbols, denoted by $\{s^p_i\}_{i=1}^{\Ntraining}$, 
over a duration of $\Ntraining \cdot \Ndata$ seconds. 



We focus on regimes in which the number of  pilot symbols $N_T$ is relatively small; and hence, it is unlikely to span a sufficient amount of different realizations of the fading coefficient at a single user. While the number of pilots is limited, we allow the users to collaborate and share their detection mappings, in order to jointly learn a unified symbol detector, i.e., one that is universally applicable not only to the users participating in the training, but also to any arbitrary user, by learning a broad fading distribution from the pilots received at all the users and the corresponding messages, i.e., $\{\mySet{D}_u\}_{u=1}^U$, where $\mySet{D}_u = \{s^p_i, \mySet{R}_i^u\}_{i=1}^{\Ntraining}$, with $\mySet{R}_i^{(u)} \triangleq \{ r_u(t) | t\in [(i-1)\Ndata, i\Ndata)\}$. 

The resulting symbol detector at an arbitrary user $u'$ recovers the message $s_{u'}$ from the received ${r}_{u'}(t)$, $t\in \Tdata+[0,\Ndata)$, assuming only the knowledge of the channel input-output relationship \eqref{eqn:ChModel}, but not the fading distribution $p_h(\cdot)$. Our proposed NN-based symbol detection scheme, which combines model knowledge of \eqref{eqn:ChModel} with data-driven tools to utilize $\{\mySet{D}_u\}_{u=1}^U$ in a collaborative fashion to train such a universal receiver, is detailed in the following section.

	\begin{figure}
	\centering
	\includegraphics[width=\linewidth]{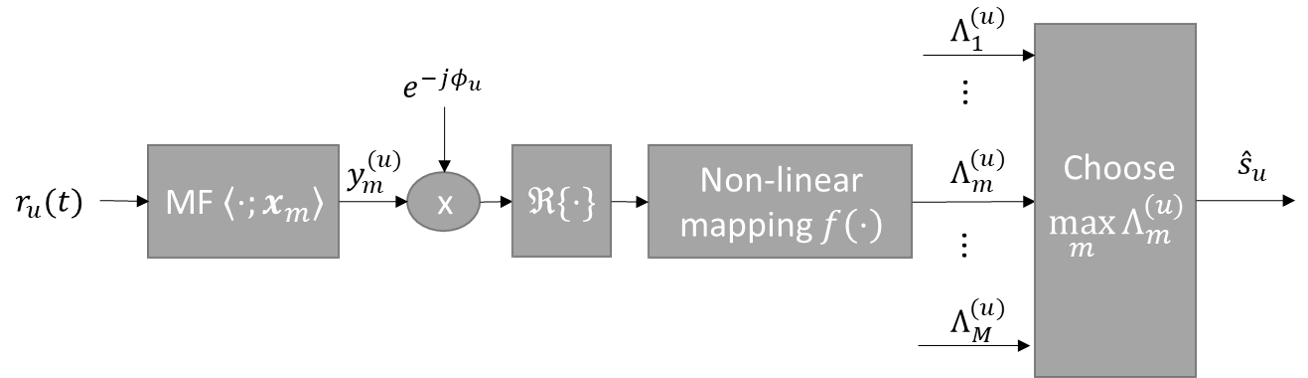} 
	\caption{\ac{map} symbol detector block diagram.} 
	\label{fig:ModelRx_1}
\end{figure}

\vspace{-0.2cm}
\section{Federated Receiver}\label{sec:UnFadeNet}
\vspace{-0.1cm}
The need for a universal symbol detector applicable to a diverse range of channel fading statistics without the exact knowledge of the underlying distribution motivates using \acp{dnn}, which were empirically shown to operate reliably in complex unknown statistical environments \cite{bengio2009learning}. However, since \acp{dnn} require large volumes of diverse training data, having each user train a local \ac{dnn} based on its limited  pilot observations is expected to yield a less reliable model with limited generalization performance. To overcome this, we design  \FadeNet~based on the following guidelines: 
(i) The fact that the channel is modeled via \eqref{eqn:ChModel} is exploited as partial domain knowledge for feature extraction, inspired by the \ac{map} rule for such channels. This approach facilitates utilizing  compact \acp{dnn} (avoiding feature extraction layers), which are trainable using relatively small datasets.  
(ii) While each local training dataset $\mySet{D}_u$ encapsulates a relatively small number of fading channel realizations, the diversity among these sets at different users is exploited to obtain a unified model for all the users by training in a federated manner.


\vspace{-0.2cm}
\subsection{Model-Based Symbol Detection for Fading Channels}
\label{subsec:ModelBasedRx}
\vspace{-0.1cm}
Here, we briefly recall the model-based \ac{map} symbol detector, which requires knowledge of $p_h(\cdot)$. We focus on scenarios in which the fading coefficient $h_u(t)$ takes a single realization during the transmission of each message, and that its phase, denoted by $\phi_u$, is known. The following derivation is based on \cite[Ch. 7.2]{simon2005digital}. 

Let $\mySet{R}^{(u)} \triangleq \{{r}_u(t)\}_{t=\Tdata}^{\Tdata+\Ndata}$. Since the message $s_u$ is uniformly distributed, the \ac{map} rule at the $u$th user is given by
\begin{equation}
\label{eqn:MAP}
\hat{s}_u^{\rm map} = \mathop{\arg \max}\limits_{m \in \CnstSet} \Pr \left(\mySet{R}^{(u)} | s_u=m \right).
\end{equation}
By defining 
    $y_{m}^{(u)} \triangleq \int _{t=\Tdata}^{\Tdata+\Ndata}r_u(t) x_m(t) dt \triangleq \langle {r}_u; {x}_m \rangle$,
  and similarly, $e_m \triangleq \langle {x}_m; {x}_m \rangle$,  the conditional distribution in \eqref{eqn:MAP} becomes
\begin{align}
\label{eqn:CondDist}
\Pr \left(\mySet{R}^{(u)} | s_u\!=\!m \right) 
&\!=\! c \int_{0}^{\infty}\! e^{\alpha \Re\{e^{-j \phi_u} y_{m}^{(u)} \}\! -\! \alpha^2e_m}p_{|h|}(\alpha) d\alpha,
\end{align}
where $c > 0 $ is a constant that does not depend on $m$ and $\mySet{R}^{(u)}$. 

The complex structure of the conditional distribution in \eqref{eqn:CondDist} for general $p_{|h|}(\cdot)$ makes evaluating \eqref{eqn:CondDist} a challenging task. If $p_{|h|}(\cdot)$ follows a simple Rayleigh distribution with scale parameter $\sigma_u$, the MAP decision criteria \eqref{eqn:CondDist} reduces to maximizing $\Lambda_m^{(u)}$ given by
\begin{align}
\label{eqn:Rayleigh}
\Lambda_m^{(u)}\!=\!\ln\{1\!+\!\sqrt{\pi}\mu_m e^{\frac{\mu_m^2}{4}}[1\!-\!Q(\mu_m/\sqrt{2})]\} \!-\!\ln(1\!+\!\gamma_m),
\end{align}
where $\gamma_m=2\sigma_u^2 e_m$, $\mu_m=\sqrt{\gamma_m/[e_m(1+\gamma_m)]} \cdot \Re\{e^{-j \phi_u} y_{m}^{(u)} \}$  and $Q$ denotes the Gaussian $Q$-function \cite[Ch. 7.2.1]{simon2005digital}. 

Despite its complex form, \eqref{eqn:CondDist} reveals that, regardless of the fading distribution, \ac{map}  detection over any fading channel conforming to \eqref{eqn:ChModel}, is comprised of the following steps: First, the observations $\mySet{R}^{(u)} $ are processed by a set of \acp{mf} to produce $\{y_m^{(u)}\}_{m=1}^{\Cnst}$; then, when the phase information is given, the \ac{mf} outputs are phase-shifted accordingly and their real part is taken; this quantity is processed by a non-linear mapping $f(\cdot)$ that depends on the fading distribution, encapsulating the integral in \eqref{eqn:CondDist}, to produce $\Lambda_m^{(u)} \propto \Pr \big(\mySet{R}^{(u)}  | s=m\big)$. This procedure is illustrated in Fig.~\ref{fig:ModelRx_1}.

	\begin{figure}
	\centering
	\includegraphics[width=\linewidth]{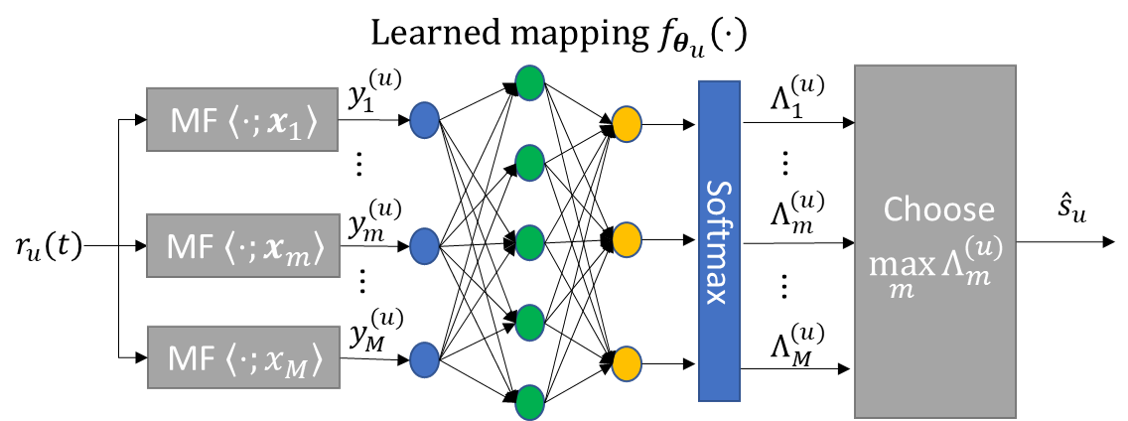} 
	\caption{The proposed symbol detector architecture.} 
	\label{fig:UnFadeNet}
\end{figure}

\vspace{-0.2cm}
\subsection{Data-Driven Symbol Detector Architecture}
\label{subsec:Architecture}
\vspace{-0.1cm}
We next present a \ac{dnn} architecture which learns to recover $s_u$ from $\mySet{R}^{(u)}$. Recall that \acp{dnn} process inputs that can be represented as vectors. Here, the input is a \ac{ct} signal $r_u(t)$, defined over the uncountable set $t\in [\Tdata,\Tdata+\Ndata)$. The intuitive approach to design the network is therefore to uniformly sample $r_u(t)$ via, e.g., Nyquist rate sampling. Taking a large number of samples is expected to facilitate handling the presence of noise and unknown channel \cite{shlezinger2020learning}, while resulting in processing high dimensional inputs, which in turn typically involves using highly-parametrized \acp{dnn}.

Nonetheless, the model-based \ac{map} detector in Fig.~\ref{fig:ModelRx_1} reveals that the \ac{mf} outputs  $\{y_m^{(u)}\}_{m=1}^{\Cnst}$ constitute a sufficient statistics for identifying the message $s_u$. As a result, we exploit this domain knowledge, and design a classification network whose inputs are the matched filter features $\{y_m^{(u)}\}_{m=1}^{\Cnst}$. This significantly simplifies the NN architecture in comparison with feeding the Nyquist samples of the received signal as input. The proposed NN architecture avoids additional layers that would be required for feature extraction from Nyquist samples of the received signal and directly uses the matched filter outputs as features for classification. 

As matched filter outputs take complex values, the \ac{dnn} input is a $2M\times1$ dimensional vector, which feeds a fully connected network. As $s_u$ can take one of $M$ different values, we use a softmax output layer with $M$ possible labels. Letting $\myVec{\theta}_u$ be the \ac{dnn} parameters  of the $u$th user, and $f_{\myVec{\theta}_u}:\mySet{C}^\Cnst\mapsto[0,1]^\Cnst$ be its mapping,  then the NN prediction can be written as
\begin{equation}
    \hat{s}_u = \mathop{\arg \max}\limits_{m\in\CnstSet}f_{\myVec{\theta}_u, m}\left(\{y_m^{(u)}\} \right),
    \label{eqn:NetOut}
\end{equation}
where $f_{\myVec{\theta}_u, m}(\cdot)$ is the $m$th output.
The  NN  is illustrated in Fig.~\ref{fig:UnFadeNet}. 

Note that while the NN architecture in Fig.~\ref{fig:UnFadeNet} is comprised of $\Cnst$ \ac{mf} components, in many transmission schemes the outputs of some of the \acp{mf} can be obtained as linear combinations of the remaining features. In particular, in modulation schemes where the modulation order is less than $M$, (e.g., 2 for QAM),  we use fewer MF features (determined by the modulation order) to simplify the NN.


\vspace{-0.2cm}
\subsection{\FadeNet~Training}
\label{subsec:Training}
\vspace{-0.1cm}
Here, we describe how the \ac{dnn} is trained. In particular, we consider a training procedure in which the users collaborate in a federated manner, exploiting the diversity in the observed channels. The resulting learned symbol detector is referred to as {\em \FadeNet}. 

As \FadeNet~is comprised of a neural classifier, we use the empirical cross entropy loss for training. The resulting loss  is 
\begin{equation}
    \mySet{L}_u(\myVec{\theta}, \mySet{D}) = -\frac{1}{|\mySet{D}|}\sum_{\{s_i,\mySet{R}_i\} \in \mySet{D}}\log f_{\myVec{\theta}, s_i}\left(\left\{ \langle \mySet{R}_i; x_m   \rangle \right\}_{m=1}^{\Cnst}  \right).
    \label{eqn:NetLoss}
\end{equation}
When the training set $\mySet{D}_u$ captures a sufficiently large number of realizations of the fading channel $h_u$, then each user should be able to tune its local \ac{dnn} parameters to carry out accurate detection. This can be achieved via conventional training mechanisms, e.g., \ac{sgd}, for which $\myVec{\theta}_u$ is iteratively updated via
\begin{equation}
    \label{eqn:SGD}
    \myVec{\theta}_u^{(n)} = \myVec{\theta}_u^{(n-1)} - \eta_n\nabla \mySet{L}_u(\myVec{\theta}_u^{(n-1)}, \{s_{i_n},\mySet{R}_{i_n}\}),
\end{equation}
where $n$ is the iteration index, $\eta_n > 0$ is the step-size, and $i_n$ is an index drawn uniformly in an i.i.d. fashion from $\{1,\ldots,\Ntraining\}$.

	\begin{algorithm} [t] 
		\caption{ \FadeNet~Training}
		\label{alg:Algo1}
		\KwData{Initial parameters $\myVec{\theta}_u^{(0)} = \myVec{\theta}^{(0)}$, $\forall u \in \NusersSet$. }
		\For{each $n = 1,2,\ldots$  }{
			{Each user $u$ sets $\myVec{\theta}_u^{(n)}$ via \eqref{eqn:SGD}}  \;
			\If{$n$ is an integer multiple of $\SGDIter$}
			{ Each user $u$ sends $\myVec{g}_u^{(n)} = \myVec{\theta}_u^{(n)}-\myVec{\theta}_u^{(n-\SGDIter)}$ to  \ac{bs}\;
			 \ac{bs} computes $\myVec{\theta}^{(n)} =\myVec{\theta}^{(n-\SGDIter)} +  \frac{1}{\Nusers}\sum\myVec{g}_u^{(n)} $ \;
			 \label{stp:DownTrans} \ac{bs} distributes $\myVec{\theta}^{(n)}$ s.t. $\myVec{\theta}_u^{(n)} = \myVec{\theta}^{(n)}$, $\forall u \in \NusersSet$\;
			}
		} 
		\KwOut{Trained \FadeNet~$\myVec{\theta}^{(n)}$ shared among all users.}
	\end{algorithm}

Nonetheless, when the channel coefficient takes a limited number of realizations in each interval of $\Ntraining\Ndata$ seconds, the local dataset may not suffice to capture the subtleties of a universal fading distribution. In such cases, the trained \ac{dnn} is likely to be highly biased towards its observed channel conditions, and may not operate reliably for future realizations of the channel in case the fading statistics change over time. To tackle this challenge, we exploit the fact that while each user may observe a limited amount of channel realizations, the overall set of different realizations observed by all the users in a cell is likely to be sufficiently diverse to train the \ac{dnn} accurately. 

Based on the above insight, we propose to jointly train a single instance of the NN shared by all the users  via \ac{fl} \cite{mcmahan2016communication}. Such distributed learning orchestrated by the \ac{bs}  requires the users to periodically exchange and synchronize their local parameters, possibly by utilizing low-rate transmissions, to train a reliable universal symbol detector. In particular, the training mechanism, based on the local \ac{sgd} algorithm \cite{stich2018local}, consists of $\SGDIter$ \ac{sgd} iterations as in \eqref{eqn:SGD}, carried out by each user locally, after which the \ac{bs} averages the trained updates  into a global parameter vector $\myVec{\theta}$, which is  distributed to  the users. This training mechanism is summarized as Algorithm~\ref{alg:Algo1}.

\vspace{-0.2cm}
\subsection{Three-Phase Operation}
\label{subsec:PhasedOperation}
\vspace{-0.1cm}
FedRec requires the users to exchange messages with the \ac{bs} during training, i.e., before the \ac{dnn} is tuned. Consequently, its application involves a three phase scheme: (i) data collection, (ii) federated training, and (iii) pilot-free communication using FedRec.

During phase (i), wireless users in the coverage area of the BS utilize a conventional pilot-based scheme to communicate with the BS. However, they keep collecting the noisy received pilot signals $\mySet{R}_i^u$ and the corresponding labels $s^p_i$ to form local datasets $\mySet{D}_u = \{s^p_i, \mySet{R}_i^u\}_{i=1}^{\Ntraining}$. To communicate with the BS during this phase, the users perform coherent symbol detection utilizing channel estimates from the received pilots.  

During phase (ii), the wireless users utilize their local datasets $\{\mySet{D}_u\}_{u=1}^U$ to collaboratively train FedRec in a federated fashion. They continue to utilize  pilot-aided coherent communications to exchange model updates with the BS for federated training.

Finally, during phase (iii), users utilize the trained FedRec receiver for pilot-free communication. As FedRec is trained with data from many users with diverse channel conditions, it does not rely on the knowledge of the exact channel realization; and hence, eliminates the need for periodic pilot transmissions. During this phase, the BS can use a low rate control channel to transmit the trained FedRec receiver to any new user entering its coverage area.

\begin{table}
\centering
\caption{BER comparison for various symbol detectors, $U=5$.}
\label{tbl:5Methods}
\begin{tabular}{|c|c|c|c|c|c|}
\hline
\multicolumn{2}{|c|}{$\rho$ (dB)} & 5 & 7.5 & 10 & 12.5 \\ \hline\hline
\multirow{5}{*}{non-iid}  & NL & 0.0677  & 0.0574 & 0.0490 & 0.0438 \\ \cline{2-6}
                          & CL & 0.0661  & 0.0544 & \textbf{0.0439} & \textbf{0.0382} \\ \cline{2-6}
                          & FedRec & \textbf{0.0649}  & \textbf{0.0530} & \textbf{0.0439} & \textbf{0.0382} \\ \cline{2-6}
                          & MD & 0.0663  & 0.0544 & 0.0445 & 0.0392 \\ \cline{2-6}
                          & MAP & \textbf{0.0647}  & \textbf{0.0526} & \textbf{0.0437} & \textbf{0.0382} \\\hline\hline
\multirow{5}{*}{iid}  & NL & 0.0577  & 0.0458 & 0.0377 & 0.0325 \\ \cline{2-6}
                      & CL & 0.0573  & 0.0453 & 0.0369 & \textbf{0.0307} \\ \cline{2-6}
                      & FedRec & \textbf{0.0561}  & \textbf{0.0444} & \textbf{0.0361} & 0.0308 \\ \cline{2-6}
                      & MD & 0.0558  & 0.0439 & 0.0357 & 0.0303 \\ \cline{2-6}
                      & MAP & \textbf{0.0557}  & \textbf{0.0439} & \textbf{0.0357} & \textbf{0.0302} \\\hline
\end{tabular}
\end{table}

\vspace{-0.2cm}
\section{Numerical Evaluations}\label{sec:sims}
\vspace{-0.1cm}
In this section we compare \FadeNet~with model-based detection and learning-based schemes for Rayleigh fading channels. We consider two cases with i.i.d. and non-i.i.d. fading channels across the users. In the i.i.d. case, the  coefficients $h_u(t)$ are generated from a Rayleigh distribution with unit scale parameter, 
and a new realization is generated on each $\Ndata$ time instances. We also consider a non-i.i.d. case, where the scale parameter for different users is not identical, e.g., due to different statistics of the local environment for each user, such that for each user $u \in \mathcal{U}$, the scale parameter $\sigma_u$ is randomized from a uniform distribution over the range $[0.5,1.5]$.

We use 16QAM modulations, i.e., $\Cnst = 16$, with average transmit power per bit $\rho$, representing the \ac{snr}. For the learning-based methods, the training and test datasets are comprised of $|\mathcal{D}_{train}|=2\cdot 10^4$ and $|\mathcal{D}_{test}|=10^7$ symbols, respectively. Training data is collected by the users during phase (i) of network operation, with each user recording $N_T=|\mathcal{D}_u|=2\cdot 10^4/\Nusers$ 16QAM pilot symbols, which are faded according to its local fading statistics. 
The learning-based methods train a NN with an input layer of size 2 corresponding to the in-phase and quadrature signal components, followed by a hidden layer with 16 neurons and softmax, thus $|\myVec{\theta}|=48$. This NN architecture is trained using the  Adam optimizer \cite{kingma2014adam} over $25$ epochs with batch-size of $20$, and tested for each \ac{snr} value. We compare the following symbol detection schemes:


\textbf{-Non-collaborative learning (NL):} Each user trains the NN solely on its local dataset. 
The \ac{ber} is averaged over  all user trained NNs using the test dataset.

\textbf{-Centralized learning (CL):} The users transmit their  datasets to the BS over a noise-free link, and the NN is trained centrally. 

\textbf{-FedRec:} The users follow Algorithm~\ref{alg:Algo1} with $5$ local epochs 
and over $5$ rounds of aggregation to train the NN collaboratively.

\textbf{-Model-based detection (MD):} The model-based detector uses the decision rule \eqref{eqn:Rayleigh}, with an estimate of the Rayleigh scale parameter, denoted by $\hat{\sigma}$, obtained from the training data. Here, each user obtains a maximum likelihood estimate of its scale parameter denoted by $\hat{\sigma}_u$ utilizing its local dataset. The overall estimate is then obtained by $\hat{\sigma}^2=1/U \sum_{u=1}^{U} \hat{\sigma}^2_u$  inserted in \eqref{eqn:Rayleigh}. 

\textbf{-MAP bound:} The MAP receiver follows the closed form \eqref{eqn:Rayleigh} for the i.i.d. case. Unlike the model-based approach which estimates the Rayleigh scale parameter from the user data, we here assume exact knowledge of the scale parameter $\sigma=1$ and insert it into \eqref{eqn:Rayleigh}. For the non-i.i.d. case, however, derivation of the MAP receiver is cumbersome and we use a numerical approach to calculate the integrals in (\ref{eqn:CondDist}) and to evaluate the BER.

In Table \ref{tbl:5Methods}, we evaluate the \ac{ber} performance of these symbol detectors for various SNRs with $\Nusers=5$ users. Among the data-driven detectors, FedRec consistently outperforms the non-collaborative scheme, while achieving similar or improved BER compared to centralized learning for both the i.i.d. and non-i.i.d cases. It is observed that the BER is significantly improved over non-collaborative learning, most notably in high \acp{snr} when users collaborate either through \ac{fl} or by exchanging their local datasets with the BS for centralized training. For the i.i.d. case, this gain is  due to the increased amount of data made available for training through collaboration. For the non-i.i.d. case, the improvement is more significant as it is not only due to increased number of data samples, but also due to the  diversity captured by these samples. 

For the i.i.d. case, FedRec approaches the performance of the model-based and MAP detectors. For the more realistic non-i.i.d. case, \FadeNet~generally outperforms the model-based approach computed with the estimated scale parameter. This indicates that while the model-based approach is sensitive to uncertainty in the scale parameter, \FadeNet~learns from data how to cope with such heterogeneous fading and performs closer to the MAP lower bound.

\begin{figure}
\centering
\includegraphics[scale=.5]{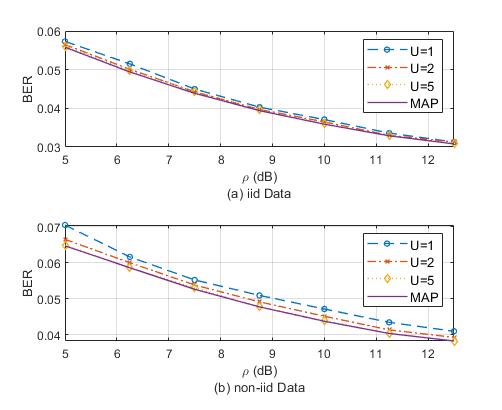}
\vspace{-0.2cm}
\caption{\FadeNet~ BER curves versus SNR.}
\label{noniid}
\end{figure}

In Fig.~\ref{noniid}, we compare the BER versus \ac{snr} of \FadeNet~for different number of users $\Nusers=1, 2, 5$. We have also added the MAP lower bound for comparison. For both i.i.d  and non-i.i.d cases, FedRec BER rapidly decreases as the number of users participating in federated training grows, approaching the MAP lower bound with merely $\Nusers=5$ users. Hence, increasing the number of users participating in federated training of \FadeNet~not only decreases $N_T$, hence reducing the duration of the data collection phase (i), but also improves the BER performance.

\begin{table}
\centering
\caption{Communication overhead in float32 words.}
\label{tbl:Overhead}
\begin{tabular}{|c|c|c|c|c|}
\hline 
\multirow{2}{*}{} & \multicolumn{2}{c|}{CL} & \multicolumn{2}{c|}{FedRec} \\ \cline{2-5} 
                  & UL              & DL             & UL               & DL            \\ \hline\hline
U=1               & 40000           & 48             & 240            & 240              \\ \hline
U=2               & 40000           & 48             & 480            & 240              \\ \hline
U=5               & 40000           & 48             & 1200           & 240              \\ \hline 
\end{tabular}
\label{tab:communications}
\end{table}

Finally, we evaluate the overhead induced on both uplink (UL) and downlink (DL) communications in the training procedure of the collaborative data-driven schemes of FedRec and centralized learning. The number of float32 words conveyed over the UL and DL channels for \FadeNet~ and centralized learning are summarized in Table~\ref{tbl:Overhead}. For centralized training, the overhead does not depend on the number of users, and is comprised of $2\cdot|\mathcal{D}_{train}|$ and $|\myVec
{\theta}|$ words on the UL and DL, respectively. For federated training, the overhead is comprised of $5$  parameter exchange rounds  of $\Nusers|\myVec{\theta}|$ words and $|\myVec{\theta}|$ words on UL and DL, respectively, both are much smaller compared to having the users transmit their  received pilots to the BS. Hence, FedRec notably reduces the communication load.

\vspace{-0.2cm}
\section{Conclusion}\label{sec:Conclusions}
\vspace{-0.1cm}
We proposed \FadeNet, a data-driven symbol detector for downlink fading channels. \FadeNet~is comprised of a \ac{dnn} designed based on the \ac{map} rule for fading channels, combined with a collaborative training mechanism, which exploits the channel diversity across multiple users through federated learning. Our numerical results demonstrate that \FadeNet~approaches the \ac{map} performance, while achieving improved robustness to uncertainty, and significantly reduces the communication overhead compared to centralized training. 


\bibliographystyle{IEEEtran}
\bibliography{IEEEabrv,refs}

\end{document}